\documentstyle[prl,aps,multicol,epsf]{revtex}
\newcommand{\beq}{\begin{equation}}
\newcommand{\eeq}{\end{equation}}

\begin{document}
\tightenlines

\title{Competition between superconductivity and the pseudogap phase
in the t-J model}

\author{{R. Zeyher}$^a$\footnote{ Corresponding author,
Phone:+49 711 689 1557, Fax:+49 711 689 1702, 
e-mail:R.Zeyher@fkf.mpg.de} 
and {A. Greco}$^{a,b}$} 

\address{$^a$ MPI-FKF, Stuttgart, Germany, \\
$^b$FCEIA and IFIR(UNR-CONICET), 2000-Rosario, Argentina}

\date{\today}


\maketitle

\begin{abstract} 
The t-J model in the large N limit (N denotes the number of spin components)
yields a pseudogap
phase in the underdoped region which is related to a d-wave charge density
wave (d-CDW). We present results for the doping dependence of
the superconducting and d-CDW order parameters as well as for collective
excitations in the presence of these two order parameters. We argue that the
electronic Raman spectrum with $B_{1g}$ symmetry probes the amplitude
fluctuations of the d-CDW at zero momentum.

\end{abstract}



The superconducting transition temperature $T_c$ in the cuprates shows
a maximum at optimal doping $\delta = \delta_c$ and a monotonic decrease
towards lower or higher dopings. On the other hand, the excitation
gap in the one-particle spectrum, often called pseudogap, increases 
monotonically with decreasing doping, remains finite above $T_c$ in the 
underdoped region, and has d-wave symmetry.
Generalizing the t-J model from 2 to N spin components it has been shown
\cite{Cappelluti} that such a pseudogap arises in the large N limit
of the t-J model
due to an instability of the normal or the superconducting state with respect 
to a d-CDW (sometimes also called flux or bond-order wave).
In the following we present results for the symmetry broken states,
namely the doping dependence of the superconducting
and the d-CDW order parameters and the dynamics of d-wave charge density
fluctuations in the presence of these order parameters. The relevance 
of a d-CDW for high-T$_c$ superconductors has recently also been 
discussed in Ref.\cite{Chakravarty}.

  In the case of the t-J model the d-CDW order parammeter $\Phi$
is defined as
\begin{equation}
\Phi = {{-2iJ}/{N_c}}\sum_{{\bf k}\sigma}\gamma({\bf k})
\langle {\tilde c}^\dagger_{{\bf k}\sigma}\tilde{c}_{{\bf k + Q}\sigma}
\rangle.
\end{equation}
$J$ is the Heisenberg coupling, $\tilde{c}^\dagger,\tilde{c}$
are creation and annihilation operators for electrons under the
constraint that double occupancies of lattice sites are excluded, $N_c$
is the number of primitive cells, $\langle ...\rangle$ 
denotes an expectation value, and $\bf Q$ is the wave vector of the d-CDW.
$\gamma(\bf k)$ is equal to $(cos(k_x)-cos(k_y))/2$.
Keeping only the instantaneous contribution in the effective interaction
\cite{Zeyher} the order parameter $\Delta$ for d-wave superconductivity is
\begin{equation}
\Delta = {2(J-V_c)/{N_c}} \sum_{\bf k} \gamma({\bf k})
\langle {\tilde c}_{{\bf k}\uparrow} 
{\tilde c}_{-{\bf k}\downarrow}
\rangle.
\end{equation}
$V_c$ is a repulsive nearest-neighbor Coulomb potential which is needed
to stabilize the d-CDW with respect to phase separation\cite{Cappelluti}.
In the presence of the above two order parameters the operators
$({\tilde c}^\dagger_{{\bf k},\uparrow},\tilde{c}_{-{\bf k},\downarrow},
{\tilde c}^\dagger_{{\bf k+Q},\uparrow},\tilde{c}_{{\bf -k-Q},\downarrow})$
are coupled leading to the following Green's function matrix\cite{Cappelluti}
\begin{equation}
G^{-1}(\omega,{\bf k}) = \left( 
\begin{array}{c c c c}  
\omega-\epsilon({\bf k})      & -\Delta({\bf k})  & -i\Phi({\bf k})                      &  0                 \nonumber\\
-\Delta({\bf k})              &\omega+\epsilon({\bf k})
&   0                         &i\Phi({\bf \bar{k}})      \\
i\Phi({\bf k})      &   0
&\omega-\epsilon({\bf\bar{k}})& -\Delta({\bf \bar{k}})  \nonumber\\
          0         &-i\Phi({\bf \bar{k}})
&-\Delta({\bf\bar{k}})&\omega+\epsilon({\bf \bar{k}})  

\end{array} \right)
\label{matrix}
\end{equation}
$\epsilon({\bf k})$ is the one-particle energy, 
$\epsilon({\bf k}) = -(\delta t +\alpha J)(cos(k_x)+cos(k_y))
-2t'\delta cos(k_x)cos(k_y)  -\mu$,
with $\alpha = 1/N_c \sum_{\bf q} cos({q_x})f(\epsilon({\bf q}))$.
$f$ is the Fermi function, $\delta$ the doping away fom half-filling,
$\mu$ a renormalized chemical potential,
$t$ and $t'$ are nearest and second-nearest
neighbor hopping amplitudes, $\omega$ a (complex) frequency, and $\bf {\bar{k}}
= {\bf k-Q}$. $\Phi({\bf k})$ and $\Delta({\bf k})$ are equal to
$\Phi \gamma({\bf k})$ and $\Delta \gamma({\bf k})$, respectively.

Expressing the expectation values in the order parameters by $G$ and
using Eq.(\ref{matrix}) one obtains coupled equations for the 
two order parameters. The thick solid and dashed lines in Fig. 1 show the 
numerically determined doping dependence
of $\Phi$ and $\Delta$ at zero temperature, calculated for $t'/t=-0.35$,
$J/t=0.3$, $V_c/t=0.06$, and ${\bf Q}=(\pi,\pi)$. The energy unit is $t$. 
The thin solid and dashed lines show $\Phi$ and 
$\Delta$ for non-interacting order parameters putting the second order 
parameter to zero. In this case the 
onset of $\Phi$ is much higher than in the interacting case, especially
\begin{figure}[h]
\centerline{
      \epsfysize=6cm
      \epsfxsize=7cm
      \epsffile{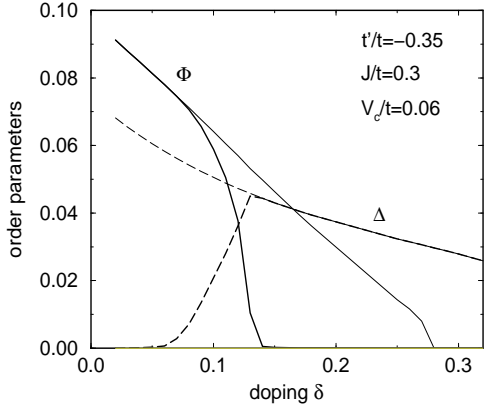}}
\label{fig1}
\caption 
{Order parameters $\Phi$ and $\Delta$
as a function of doping in units of $t$ at $T=0$ in the interacting (thick
lines) and non-interacting (thin lines) case.}
\end{figure}
\noindent
if $t'$ 
is nonzero. This reduction in the onset of $\Phi$ reflects the competition
of the two order parameters. The superconducting order parameter
$\Delta$ increases monotonically with decreasing doping in the non-interacting
case but is heavily suppressed by $\Phi$ below the onset of $\Phi$ in the 
interacting case. For $0.05 < \delta <0.15$ both order parameters are non-zero
so that superconductivity and d-CDW coexist throughout this region.

The eigenvalues of the matrix $G^{-1}$ give the one-particle energies in
the presence of the two order parameters. The corresponding density of states
$\rho(\omega)$ is shown in Fig.2 for three different dopings. 
In the upper diagram, corresponding to the overdoped case with $\Phi = 0$,
the usual density for a d-wave superconductor is seen. The lowest
diagram in Fig. 2 corresponds to the underdoped region dominated by
the d-CDW. The zero in $\rho(\omega)$ occurs somewhat above the 
chemical potential $\omega = 0$ and $\rho(\omega)$ is rather asymmetric
around this point. The diagram in the middle of Fig.2 describes a slightly
underdoped case where both order parameters are of similar magnitude.
$\rho(\omega)$ reflects here both gaps and possesses additional structures
due to the geometry of the two-dimensional Fermi surface and the Umklapp
processes induced by $\bf Q$.
\begin{figure}[h]
\centerline{
      \epsfysize=6cm
      \epsfxsize=7cm
      \epsffile{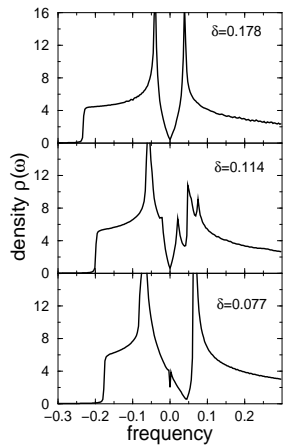}}
\label{fig2}
\caption 
{One-particle density of states for three different dopings.}
\end{figure}

The above calculations were performed assuming a commensurate 
${\bf Q}= (\pi,\pi)$. In reality the d-CDW is at low temperatures and a 
finite doping incommensurate\cite{Cappelluti} with four inequivalent
wave vectors ${\bf Q}=\pm(\pi,q),\pm(q,\pi)$ with $q$ different from $\pi$.
Writing $q=\pi-x$ the incommensuration $x$ was approximately calculated 
from the leading instability in ${\bf k}$-space using a linear approximation,
$x$ is shown in Fig. 3 as a function of $\delta$.
The left diagram in this figure corresponds
to $\Delta=0$, i.e., the thin line in Fig. 1. In this case $x$ assumes the
rather large value of about 0.6 at the onset of the d-CDW, but then decays
rapidly to about 0.1 with decreasing doping. The right diagram in Fig. 3
illustrates that superconductivity not only depresses the onset of the
d-CDW but also the incommensuration $x$ resulting in a practically
commensurate d-CDW in the underdoped regime.
In our calculations $x$ has been determined from that momentum where the 
d-wave charge susceptibility in the superconducting or normal state 
shows the strongest divergence.
\begin{figure}[h]
      \centerline{\hbox{
      \epsfysize=6cm
      \epsfxsize=6cm
      \epsffile{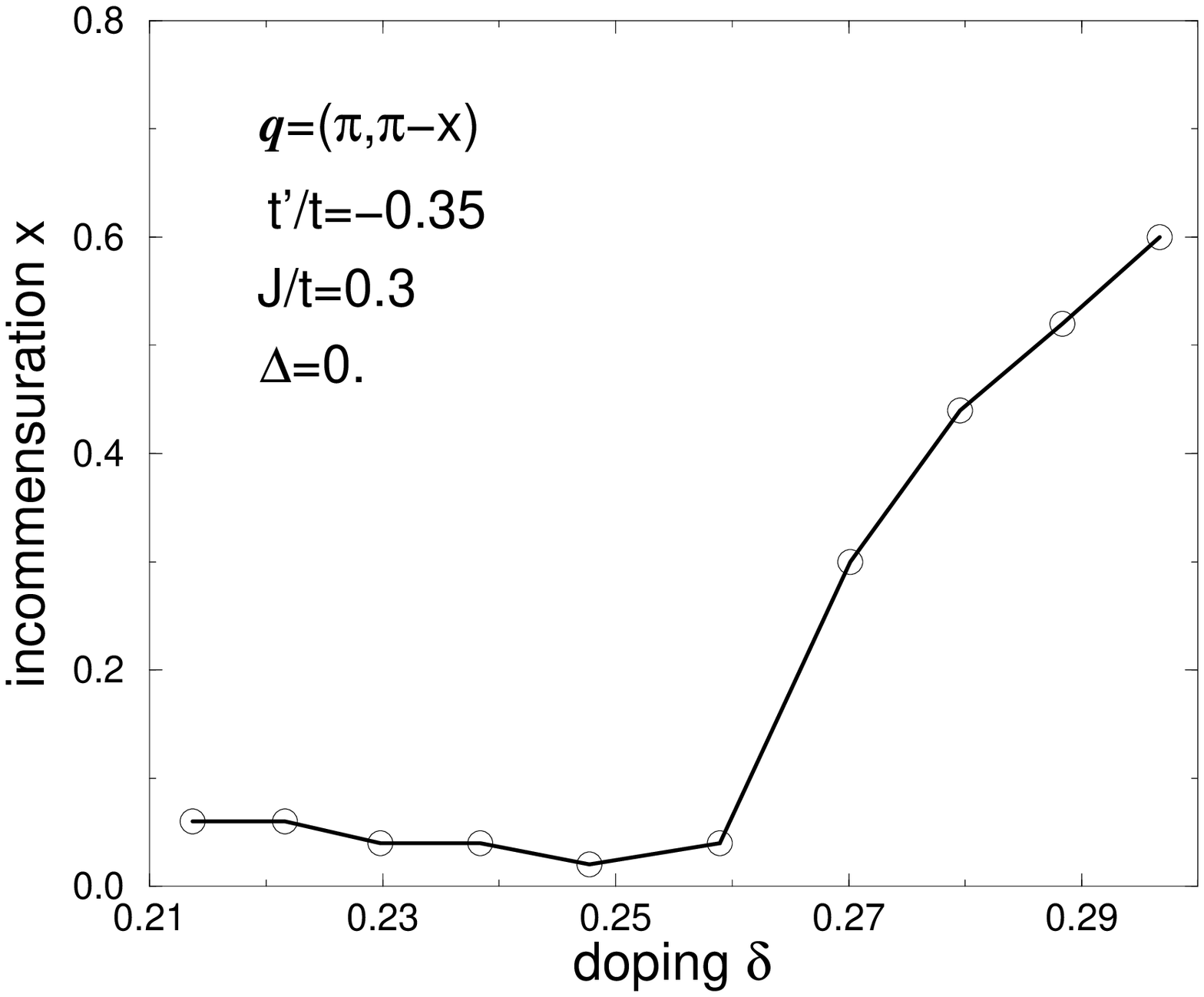}
      \epsfysize=6cm
      \epsfxsize=6cm
      \epsffile{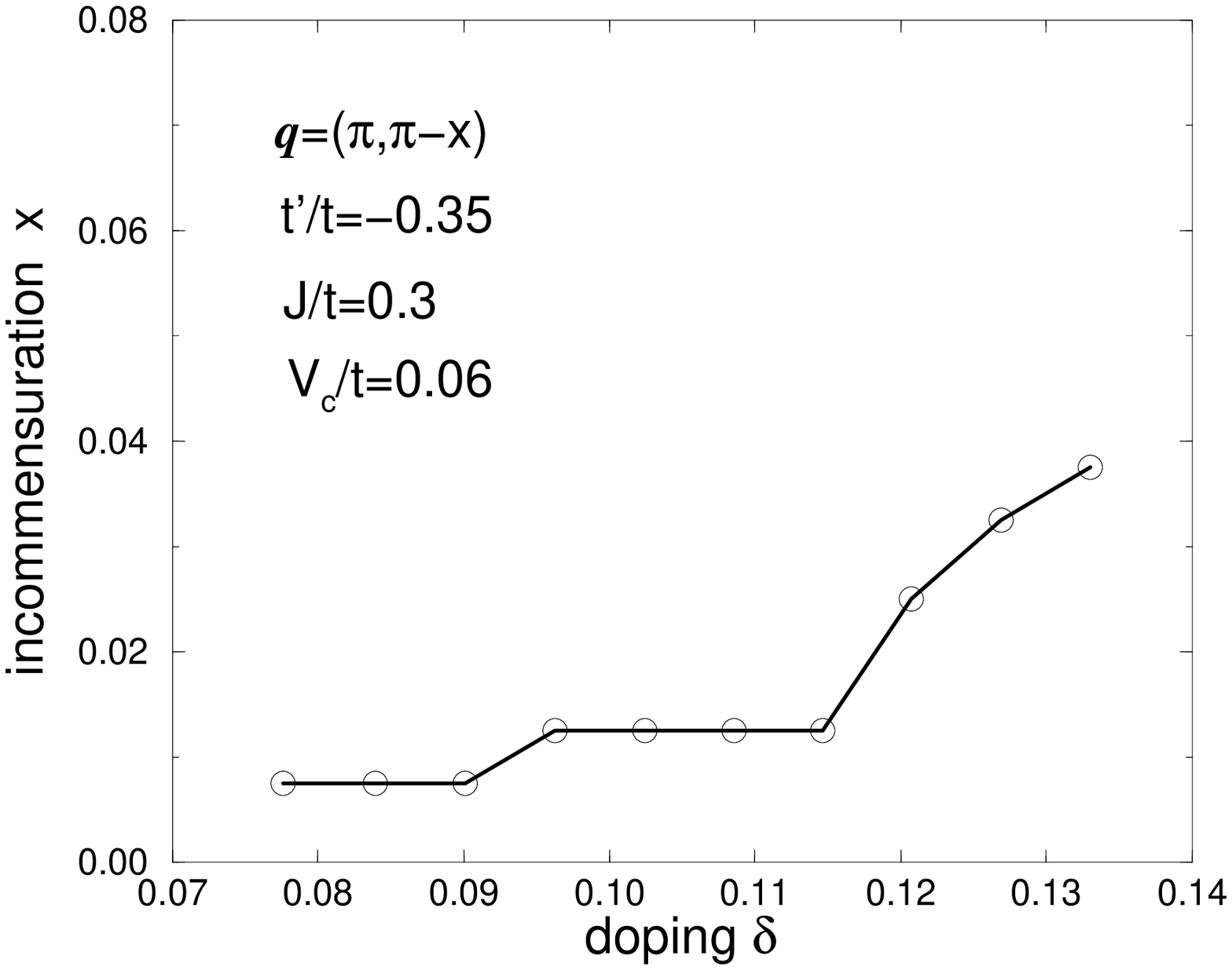}}}
\caption 
{Deviation x of one wave vector component of the d-CDW from $\pi$ as a 
function of doping in the absence (left diagram) and presence (right diagram)
of superconductivity.}
\label{fig5}
\end{figure}

Let us denote by $\chi({\bf q},\omega)$ the response function associated
with the d-wave density operator $\rho({\bf q}) = 1/N_c \sum_{{\bf k}\sigma}
\gamma({\bf k}) {\tilde c}^\dagger_{{\bf k+q},\sigma}
{\tilde c}_{{\bf k} \sigma}$. At large N $\chi$ is given by
$\chi({\bf q},\omega) = \chi^{(0)}({\bf q},\omega)/(1+J\chi^{(0)}
({\bf q},\omega))$ which holds exactly at ${\bf q}= (0,0)$ and $(\pi,\pi)$ 
and in a good approximation for a general ${\bf q}$. For a general $\bf q$
the explicit expression for $\chi^{(0)}$ is rather involved, so we give it here
only for the special case ${\bf q}=(0,0)$,
\begin{equation}
\chi^{(0)}(\omega) = P_{11,11}(\omega) -
P_{12,21}(\omega) -P_{13,31}(\omega) +P_{14,41}(\omega),
\label{chi}
\end{equation}
\begin{equation}
P_{ij,kl}(\omega) = {2T \over N_c}\sum_{{\bf k},n} \gamma^2({\bf k})
G_{ij}(i\omega_n +\omega,{\bf k}) G_{kl}({i\omega_n},{\bf k}).
\end{equation}
The left diagram in Fig. 4 shows the negative imaginary part of 
$\chi^{(0)}$ (dashed curves)
and of $\chi$ (solid curves) at ${\bf q}=(0,0)$ and three different
dopings. The curves illustrate that the Heisenberg term as the residual 
interaction
strongly scatters the quasi-particle excitations across the pseudogap
shifting most of the spectral weight down into bound states inside the gap.
In the overdoped case (upper diagram in Fig. 4) this bound state describes
an exciton state inside the superconducting gap. In the strongly underdoped 
case (lowest diagram) the bound state corresponds to the amplitude mode
of the d-CDW probed at a 
\begin{figure}[h]
      \centerline{\hbox{
      \epsfysize=6cm
      \epsfxsize=7cm
      \epsffile{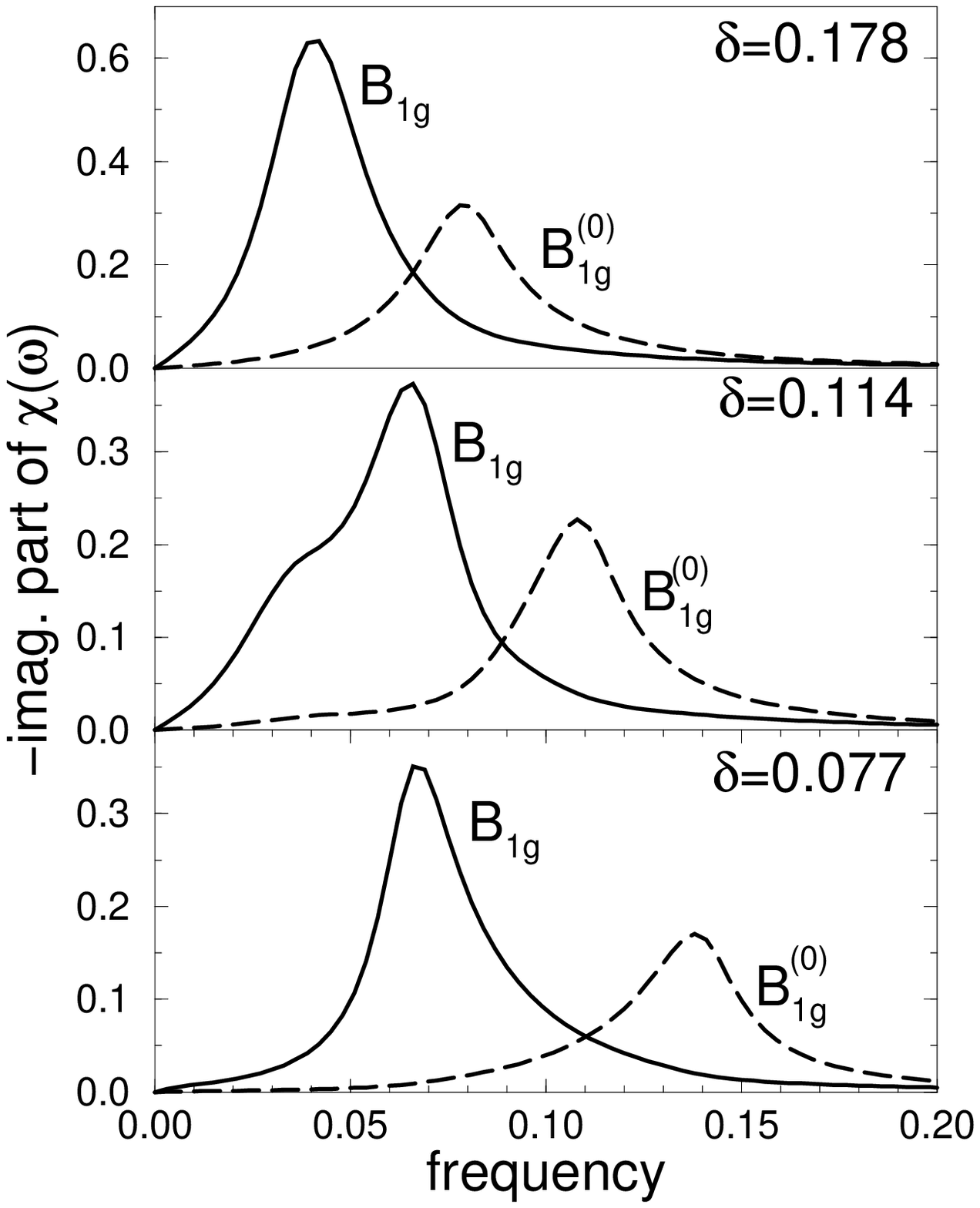}
      \epsfysize=6.75cm
      \epsfxsize=6cm
      \epsffile{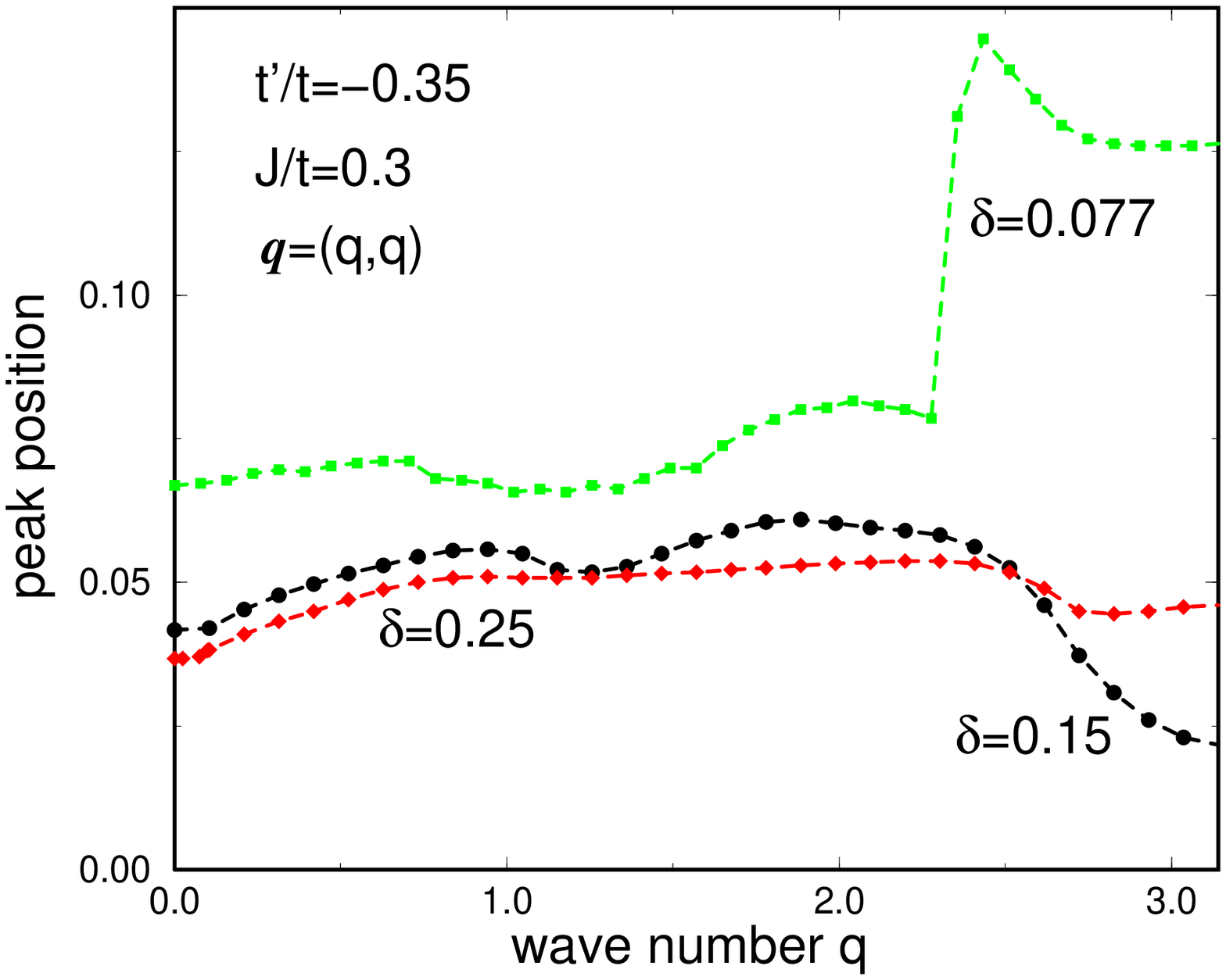}}}
\caption 
{Left diagram: Correlation function for d-wave density fluctuations at 
${\bf q}=(0,0)$ in the free (dashed lines, called $B^{(0)}_{1g}$) and 
interacting (solid lines, called $B_{1g}$) case; Right diagram: dispersion
of the main peak in the d-wave density fluctuation spectrum  along $(q,q)$ 
for three different dopings.}
\end{figure}
\noindent
wave vector $(\pi,\pi)$ away from the
wave vector of the d-CDW, i.e., at ${\bf q}= (0,0)$. With decreasing doping the
frequency of the peak increases monotonically and thus is tight to the
total pseudogap. The solid curves in Fig.3 are
proportional to the $B_{1g}$ spectra of electronic Raman scattering
in the cuprates and are in good agreement with the experimental 
data\cite{Kendziora}. In particular, one can understand why
the $B_{1g}$ peak does not probe the superconducting but the 
pseudogap\cite{Zeyher1}.

In the right-hand diagram of Fig.4 we plotted the position of the main 
peak in the d-wave fluctuation spectrum along ${\bf q}=(q,q)$ for three  
different dopings. In the overdoped ($\delta =0.25$) and the slightly
overdoped ($\delta=0.15$) cases only one well-pronounced
peak in the spectrum was obtained describing collective d-wave density 
fluctuations.
Its dispersion is rather weak away from $(\pi,\pi)$. Near $(\pi,\pi)$
a well-pronounced soft mode develops if the doping approaches the onset of
the d-CDW from above. Below the onset the soft mode hardens with 
decreasing doping and its peak position is near or somewhat below
the value $2\Phi$ in analogy with BCS-theory. In the strongly underdoped 
case with $\delta=0.077$ the momentum region around $(\pi,\pi)$, where the
peak position is roughtly given by $2\Phi$, is rather small. In the 
neighborhood of $q \sim 2$ the fluctuation spectrum develops two peaks
and the spectral weight shifts from the upper to the lower peak
with decreasing momentum. The lower peak shows only little dispersion
towards smaller momenta and lies somewhat above the corresponding peaks
at larger dopings.

  In conclusion, we found that the underdoped regime of the t-J model at
large N is characterized by the competition between d-wave superconductivity
and a d-wave CDW. This competition is especially strong for a finite $t'$:
For instance, for $t'/t=-0.35$ the d-CDW onset shifts from about 
$\delta = 0.29$ to $\delta = 0.14$ due to superconductivity.
The wave vector of the d-CDW is very close to the commensurate
value $(\pi,\pi)$, except near its onset. The d-wave density fluctuation
spectrum is determined by collective effects. It exhibits a pronounced
soft mode behavior near the onset of the CDW but shows only little
dispersion far away from $(\pi,\pi)$. The fluctuation spectrum at zero 
momentum agrees well with data from electronic Raman scattering, in
particular, with respect to its dependence on doping.

The authors thank Secyt and the BMBF (Project ARG 99/007) for financial 
support and P. Horsch for a critical reading of the manuscript.

\end{document}